\documentclass[10pt,preprint]{aastex}
\usepackage{epsfig}
\usepackage{amsmath}
\usepackage{multirow}
\usepackage{apjfonts}

\def\eps@scaling{1.0}%
\newcommand\plotsix[7]{{%
\typeout{Plotsix included the files #1 #2 #3 #4 #5 #6 #7}
\centering
\leavevmode
\columnwidth=.45\columnwidth
\includegraphics*[width={\eps@scaling\columnwidth}]{#1}%
\hfil
\includegraphics*[width={\eps@scaling\columnwidth}]{#2}%
\hfil
\includegraphics*[width={\eps@scaling\columnwidth}]{#3}%
\hfil
\includegraphics*[width={\eps@scaling\columnwidth}]{#4}%
\hfil
\includegraphics*[width={\eps@scaling\columnwidth}]{#5}%
\hfil
\includegraphics*[width={\eps@scaling\columnwidth}]{#6}%
\vfill
\includegraphics*[width=15cm, height=2.2cm]{#7}%
}}%

\begin{document}

\title{On the excess of ultra-high energy cosmic rays in the direction of Centaurus~A}

\author{Ruo-Yu Liu\altaffilmark{1,2} , Xiang-Yu Wang\altaffilmark{1,2} , Wei Wang\altaffilmark{3} and Andrew M. Taylor\altaffilmark{4}}
\altaffiltext{1}{Department of Astronomy, Nanjing University,
Nanjing, 210093, China; ryliu@nju.edu.cn; xywang@nju.edu.cn} \altaffiltext{2}{Key laboratory of Modern
Astronomy and Astrophysics (Nanjing University), Ministry of
Education, Nanjing 210093, China} \altaffiltext{3}{National Astronomical
Observatories, Chinese Academy of Sciences, 20A Datun Road, Chaoyang District, Beijing 100012, China}
\altaffiltext{4}{ISDC, Chemin d'Ecogia 16, Versoix, CH-1290, University of Geneva, Switzerland}

\begin{abstract}
A posteriori anisotropy study of ultra-high energy cosmic 
rays (UHECRs) with the Pierre Auger Observatory (PAO) has shown 
evidence of excess of cosmic ray particles above 55~EeV within
$18^{\circ}$ of the direction of the radio galaxy Centaurus~A.
However, the origin of the excess remains elusive. We simulate the
propagation of different species of particles coming from the
direction of Centaurus~A in the Galactic magnetic fields, and find
that only particles of nuclear charge $Z\la 10$ can avoid being
deflected outside of the $18^{\circ}$ window of Centaurus~A. On the
other hand, considering the increasingly heavy composition of UHECRs
at the highest energies measured by PAO, a plausible scenario for
cosmic rays from the direction of Centaurus~A can be found if they
consist of intermediate--mass nuclei. The chemical composition of
cosmic rays can be further constrained by lower-energy cosmic rays
of the same rigidity. We find that cosmic ray acceleration in the
lobes of Centaurus~A is not favored, while acceleration in the
stellar winds that are rich in intermediate-mass nuclei, could meet
the requirement. This suggests that the observed excess may
originate from cosmic ray accelerators induced by stellar explosions
in the star-forming regions of Centaurus~A and/or the Centaurus
cluster located behind Centaurus~A.
\end{abstract}


\section{Introduction}
Despite nearly one--century of effort, the origin of ultra--high
energy cosmic rays (UHECRs) still  remains unknown. It is believed
that the detections of anisotropy in the arrival direction of UHECRs
can provide a useful clue for recognising the sources. The Pierre
Auger Observatory (PAO) has detected 69 comic rays with energy above
55~EeV between January 2004 and December 2009. The correlating
fraction of the arrival directions which are closer than
$3.1^{\circ}$ from the position of an AGN within 75~Mpc is $38\%$
(i.e. 27 out 69 events), while the expected fraction is $21\%$ for
an isotropic distribution of these cosmic rays.  The significance of
the anisotropy has decreased compared to previous estimates with
smaller datasets \citep{PAO07,PAO08,PAO10}. On the other hand, it is
reported that 13 events are located within $18^{\circ}$ of
Centaurus~A (Cen~A) out of a total of 69 events in the whole sky
above 55~EeV, whereas 3.2 events would be expected from this region
\citep{PAO10}, implying that the clustering is not a statistical
accident.

It has long been proposed that AGNs could accelerate particles to
ultra-high energies (e.g. \citealt{Mannheim95, Boldt99,
Berezinsky06, Dermer09}). Particularly, Cen~A, as the nearest AGN
($d\sim 3.8$~Mpc, \citealt{Harris10}), was suggested by various
authors  as one potential source of UHECRs (e.g. \citealt{Cavallo78,
Romero96, Anchordoqui01, Hardcastle09, Gureev10, Pe'er11}). However,
there are two  issues remaining to be solved before concluding that
these UHECRs are from nearby AGNs. First, most of the nearby
correlating AGNs are subluminous, so they may not be powerful enough
to accelerate protons to $10^{20}$eV (see e.g. \citealt{Waxman95,
Piran10, Waxman11}, but see \citealt{Pe'er11}), requiring
intermediate mass nuclei or heavy nuclei as the composition of
UHECRs in order to reach these energies. The other problem is the
influence of magnetic field on the trajectories of cosmic rays. Both
extragalactic and Galactic magnetic fields (EGMF and GMF) can
deflect UHECRs, preventing them from pointing back to their birth
places when observed from Earth. So the apparent arrival direction
is not necessarily the real direction of the source and this
influence is more serious for heavier nuclei. Recent measurements of
the maximum air shower elongations ${\rm <X_{max}>}$ and their
fluctuations ${\rm RMS(<X_{max}>)}$ by the PAO indicate that the
cosmic ray spectrum is { gradually dominated by intermediate or 
heavy mass nuclei up to $E \sim 4\times 10^{19}$eV}\footnote{Due to lack of statistics, all the events with energy larger than $\sim 4\times 10^{19}$eV are used in the last bin to get the composition. So we can not get any information of $X_{\rm max}$ and $RMS(<X_{\rm max}>)$ at the highest energy (i.e. $\leq 10^{20}$eV). In this paper, we assume the tendency which is shown in the lower energy range continues to the highest energy.}\citep{PAO10b}, although this claim depends on the hadronic
interaction model at high energies which is not well-known at
present {\footnote{It should be noted here that observation of High
Resolution Fly's Eye Experiment (HiRes)  did not show any
correlation between arrival direction of UHECRs and nearby AGNs, and
observation of  Telescope Array (TA) can not distinguish an
isotropic distribution from an anisotropic distribution of UHECRs
arrival directions in current statistics \citep{Abbasi04,
Tsunesada11}. Additionally, both of their elongation measurements
prefer a pure proton composition in high energy
end\citep{Abbasi10,Tsunesada11}. Our paper is based on the
observation of the PAO.}}. Interestingly, Hooper \& Taylor (2010)
find that good fits to the  ${\rm <X_{max}>}$ and  ${\rm
RMS(<X_{max}>)}$  data can be found for the case in which the
sources accelerate primarily intermediate mass nuclei.

In this paper, we examine the effect of Galactic magnetic fields on
the deflection of UHECRs  of different species that originate from
the direction of Cen~A and study the constraints on the composition
of the UHECRs imposed by the excess in the direction from Cen~A.
Since the EGMF is poorly known, its impacts on UHECR deflection is
not well understood. However, in general it is found that the deflections
by the EGMF tend to be along and within the cosmic large scale structure of
the galaxy distribution \citep{Giacinti10}. For these reasons, we
neglect the effect of the EGMF in this work and only consider the effect
of the GMF on the deflection of UHECRs.

The rest of the paper is organized as follows. In \S 2, we
briefly introduce the GMF models employed in this paper, as well as our
method for simulating the propagation of UHECRs. We show our results
and their implications in \S 3. In \S 4, we further constrain the
chemical composition of UHECRs  based on our analysis of the
lower-energy cosmic rays with the same rigidity. We discuss the
possible sources of the excess and give our
conclusion in \S 5. Throughout the paper, we use eV as the unit of
particle energy and use c.g.s units for other quantities and denote
by $Q_x$ the value of the quantity $Q$ in units of $10^x$, unless
specified otherwise.

\section{Simulating propagation of UHECRs in the GMF}

{ The Galactic magnetic fields are generally described to be composed
of  a regular component in the disk and a large-scale field in the
halo. The previous models for the magnetic field  in the disk are
either axissymmetric \citep[e.g.][]{Stanev97} or bi-symmetric
\citep[e.g.][]{Han99,Tinyakov02}. {However, neither of these simple
models agrees very well with observations \citep{Han06}.} Thus, in
the present work, we employ two updated models, one described in
\citet{Giacinti10}, which is based on the model built by
\citet{PS03} (hereafter the PS model), and another developed in
\citet{Jiang10} (hereafter the J model) respectively. Both of these
models contain a disk component and a halo component which consists
of a toroidal field and a poloidal field. The halo component in
these two models are basically the same. The field configurations
are based on the anti-symmetric rotation measure (RM) sky revealed
by the extragalactic radio sources \citep{Han97, Han99} and the
vertical filaments in the Galactic center \citep{YZ84, YZ04}, while
their strengths are selected to meet the features of the observed
filaments in the Galactic center \citep{Morris96} and the vertical
field component in the vicinity of the Sun \citep{Han94}. For the
disk component, there is a difference between these two models. In the
PS model, the disk component is developed from a basic conservative
model constructed by \citet{Han94}, which was based on the
Faraday-rotation measurement of 134 pulsars, with the assumption of
two logarithmic spiral arms and a constant pitch angle, showing
bisymmetric (BSS) magnetic field (see \citealt{PS03}). While in the
J model, the disk component is founded upon the RM of pulsars by
\citet{Han06}, and adopts a four-arm spiral structure, whose
configuration can be depicted by an Archimedes spiral \citep{Hou09},
with the magnetic fields reversed from arms to inter--arms
\citep{Jiang10,Nota10}. We present the configurations of the disk
magnetic field of these two models in Figure~1 to show the
difference. A detailed description of both the PS model and the J
model can be found in \citet{Giacinti10} and \citet{Jiang10}
respectively.}

Since the deflections of UHECRs caused by the small scale random
component at the  energies we consider are smaller or at most
comparable to that induced by the regular GMF even for iron nuclei
\citep{Giacinti10, Tinyakov05}, the random component can be ignored
as far as only the largest deflection angle is concerned in the
present study.

The propagation of cosmic ray particles can be described by
\begin{equation}
\left\{
\begin{array}{ll}
d\textit{\textbf{p}}/dt=Ze\textit{\textbf{v}}\times \textit{\textbf{B}}\\
d\textit{\textbf{x}}/dt=\textit{\textbf{v}}
\end{array}
\right.
\end{equation}
where $e$ is the charge of the electron and $c$ is the speed of
light in the vacuum. \textit{\textbf{p}},  \textit{\textbf{v}} and
\textit{\textbf{x}} are the momentum, velocity and the spatial
coordinates of the particle respectively, and \textit{\textbf{B}} is
the magnetic field. Here we do not consider any changes of the
particle during propagation, such as fragmentation or cooling
processes. Following the method used in previous works
\citep[e.g.][]{Stanev97, Harari99, Tinyakov02, PS03, Giacinti10,
Jiang10}, we propagate an anti-proton or anti-nucleus from the earth
along certain direction, tracking it until it reaches the
border of the Galaxy, { and recording its position and velocity
direction at that moment. A proton or a nucleus
which enters the Galaxy at the same position and with the same
direction of its velocity will travel back to the Earth along the
same path.} The spatial extent of cosmic rays emitted by Cen~A is
assumed to be disk-like, with a typical radius of $5^{\circ}$ on the
celestial sphere. Thus, particles which enter the Galaxy with a
direction closer than $5^{\circ}$ from the center of an object are
regarded as coming from that object.

{To study the influence of the GMF on the arrival directions of
cosmic rays coming from the direction of Cen~A,  we {
isotropically } propagate and backtrace $10^6$ particles with
energies from $10^{19.75}$eV ($\sim$ 55~EeV,  the threshold energy
for the clustering) to $10^{20.15}$eV ($\sim$ 142~EeV, the highest
energy the PAO ever detected)  with an increment of $10^{0.05}$ in
energy for different species of particles separately. There are two 
important parameters for each particle. One is the angle between the
particle's arrival direction and the direction of Cen~A, which we
denote  as $\theta$. The other one is the angle between the
direction of the particle's velocity when it enters the Galaxy and
the direction of Cen~A, denoted by $\alpha$.} Figure 2 and 3 show
the Galactic coordinates of arrival directions of particles coming
from Cen~A (i.e. $\alpha < 5^{\circ}$) for the two GMF models
respectively. Each dot represents a single particle, and the
different colors represent their energies. The black filled star is
the location of Cen~A ($l=309.52^{\circ}, b=19.42^{\circ}$) and the
black solid curve is the projection on the celestial sphere of a
circle centered on Cen~A with a radius of $18^{\circ}$. The small
open circles represent the observed events and the size of the
circles is proportional to their energies. The 13 { magenta} circles
represent the events correlated with Cen~A. One can see that, in
both GMF models, as expected, the smaller the particle's rigidity
(i.e. $E/Z$, where $E$ and $Z$ are the energy and the nuclear charge
of the particle respectively), the larger the deviation from the
original direction. For heavy nuclei, such as iron  or calcium
nuclei, the particles are severely deflected even at the highest
energy, while for lighter nuclei such as proton or helium nuclei,
the particles can reach the earth without being much deflected.

\section{Effects of GMF on the arrival directions of UHECRs and its implications}
The observed 13 cosmic ray particles from the direction of Cen~A
spans the energy range from $10^{19.75}$eV to $10^{19.95}$eV. The
influence of GMF on the arrival directions of  cosmic ray particles
of different energies can be studied quantitatively if their energy
spectrum is given. We assume that these particles are accelerated
in Cen~A to a power--law distribution with index $s$, and keep
the same spectrum whilst propagating to the border of the Galaxy, since the
distance of Cen~A to us is much shorter than the attenuation length.
Through our simulation of the propagation of these particles in the GMF, we get
a mock distribution of arrival directions for particles of
different species and different energies. For  convenient
comparisons with observations,  we define two parameters: one is
$\zeta_i(E)$, defined as the number ratio of  particles of species
$i$ and energy $E$ that reach the Earth when the deflection by the GMF
is considered  to those that reach the Earth without considering the
GMF deflection. This factor  is also known as the magnetic lensing
amplification factor \citep[see,
e.g.][]{Harari00,Harari02,Giacinti10}.  Another parameter is
$\eta_i(E)$, defined as the number ratio of   particles that reach
the Earth from a direction within $18^{\circ}$ of Cen~A (i.e. number
of particles with $\theta < 18^{\circ}$) to those that reach the
Earth from all directions. In the calculations, we consider the
relative exposure of the PAO \citep{Sommers01}. The corresponding ratios
that are integrated over energy from $10^{19.75}$eV to
$10^{19.95}$eV (approximately the energy range of the observed 13
events) can be obtained by
\begin{equation}
\bar\zeta_i=\frac{\int \zeta_i(E)\frac{dN_i}{dE} dE}{\int
\frac{dN_i}{dE}dE}
\end{equation}
and
\begin{equation}
\bar\eta_i=\frac{\int \zeta_i(E)\eta_i(E)\frac{dN_i}{dE}dE}{\int
\zeta_i(E)\frac{dN_i}{dE} dE}.
\end{equation}
Table 1 lists the values of these two ratios for different particle species
for both the PS model and the J model. From the values
of $\bar\zeta_i$, one can find that the flux of light particles are
barely influenced by the GMF due to their high rigidity. Here, the flux of
intermediate mass nuclei such as oxygen and silicon are magnified
while that of heavy nuclei are slightly demagnified. The values of
$\bar\eta_i$ are smaller as the nuclei get heavier, implying that
heavy nuclei will be deflected severely from their original
directions by the GMF.

Take iron nuclei as an example case. If the excess events  from
Cen~A are dominated by iron nuclei, the all-sky  event number of
iron nuclei contributed by Cen~A should be $13/\bar\eta_{\rm
Fe}=1300$ for the PS model, far more than the observed event number
which is 64 in the same energy range. On the other hand, it has been
shown that it is highly unlikely for all the observed events with
energy $>55$~EeV to come solely from a single source, because 
simulations of the arrival directions for such a case indicate 
that it would display an apparently dipolar pattern,
which is different from the observed distribution even in the most
favorable cases \citep{Giacinti11}. Thus, we draw a circle  centered
at Cen~A with a radius equal to the largest deflection angle (the
pale solid circle in Figure 2 and 3), and regard the observed event
number in the same energy range that the circle covers as the
maximum event number ($N_{{\rm max},i}$) that Cen~A can contribute
to. For instance, for iron nuclei in the PS model, the maximum event
number is 33.  In some cases (e.g. for light nuclei), the pale
circles shrink into the $18^{\circ}$ window of Cen~A, and we set
$N_{\rm max}$ to be 13 in such cases. For iron nuclei in the PS
model, they could contribute at most $N_{\rm CenA, Fe}=N_{\rm max,
Fe}\times \bar\eta_{\rm Fe}=0.33$  events in the $18^{\circ}$ window
of Cen~A. Thus, as mentioned above, we disfavor iron nuclei as the
dominant composition of the excess events.

Similarly, we also calculate the maximum event number that other
species of particles can contribute for both GMF models. The results
are shown in Table 1. One can see that the observed excess events
around Cen~A can be reconciled to particles with $Z\la 10$ in the
both models. Since heavy nuclei such as
iron will be deflected severely, deviating far from their original
direction, they can be ruled out as the main composition of the
excess events from Cen~A. We note that the distribution of simulated
particles within the $18^{\circ}$ window is concentrated in some
regions when only the regular component of the GMF is considered,
which is not as scattered as the observed distribution.  Since we
aim to obtain constraints on the composition of excess cosmic rays,
for which only the largest deflection angle is concerned,
modeling the { actual} distribution is beyond the present work. 
{ However, despite this, we do note that the incorporation of random 
components of the GMF and/or appropriate EGMF would likely wash out the 
distribution obtained in the present simulation, making it look more 
like the observed event distribution} \citep[e.g.][]{Yuksel12}.

\section{Constraints on the source composition with lower-energy cosmic rays}
The shower profile of UHECRs measured by PAO implies that the
average chemical composition of cosmic rays consists of heavy or
intermediate--mass nuclei, most notably at the highest energy bin,
as long as the properties of hadronic interactions do not change
significantly at such high energies. If the measured
all-sky--averaged composition roughly reflects the average
composition of particles from the direction of Cen~A, light
particles, such as protons and helium nuclei, can not be the
dominant component of these events. Such a heavy  or
intermediate--mass nuclei composition could be achieved only if 1)
the accelerated material is rich in intermediate--mass or heavy
elements; or 2) the acceleration ability of the source is limited so
that only intermediate-mass or heavy nuclei can be accelerated to
energies $>10^{19.75}$ ~EeV (i.e., the source is not powerful enough
to accelerate light particles to energies $>10^{19.75}$ ~EeV). In
the latter scenario, the composition in the energy range from
$10^{19.75}$eV to $10^{19.95}$eV is dominated by intermediate--mass
nuclei or heavy nuclei, regardless of the chemical composition of
the accelerated material. However, one should be cautious of
lower-energy lighter cosmic rays that have the same rigidity, as
they will follow the same trajectories and could produce stronger
anisotropy in the direction of Cen~A at lower energies if their
abundance is not low \citep{Lemoine09,PAO11}. We will  show below
that the chemical composition of the accelerated material in this
scenario must satisfy some constraint as well.

A number of acceleration sites of Cen~A, such as inner jets/lobes
\citep[see, e.g.][]{Dermer09, Rieger09,Honda09}, the north middle
lobe (NML) \citep[see, e.g.][]{Romero96} as well as the giant lobes
\citep[see, e.g.][]{Hardcastle09,OS09}, have been suggested for
UHECRs\footnote{The central black hole with a strong magnetic field
has also been suggested as a possible accelerator
\citep[e.g.][]{Neronov09}, though other works have highlighted that
the maximum energy may be somewhat limited, even for the of case of
nuclei, see e.g. \citep{Lemoine09, Rieger09}. For this reason we do
not consider it in the present work.}. { Although there are no
precise measurements of the element abundance of these possible
acceleration regions in Cen~A, we can roughly estimate the abundance
from various observations  in these regions.} In the inner lobe
case,  UHECRs could be accelerated by shocks driven by the inner
lobes inflating into the interstellar medium (ISM). So the chemical
composition of the accelerated UHECRs in this case should be similar
to that of the ISM. Deep Chandra observations of some nearby
gas--rich elliptical galaxies show a near--solar metallicity of ISM
generally, except an unequivocally sub--solar oxygen abundance ($\la
0.5Z_{\odot}$, see \citealt[][and reference therein]{Kim12}). For
the NML, the spectra of five associated X--ray knots can be well
fitted by sub--solar element abundance \citep{Kraft09}, implying a
sub--solar metallicity for NML. As for the giant lobes, since they
extend out to 600~kpc from the center, the ambient medium should be
intergalactic medium (IGM), so the chemical composition of
accelerated UHECRs is probably identical to that of IGM, which is
found to be sub--solar metallicity both observationally and
theoretically \citep[e.g.][]{Richter09, Barai11}. Here we assume the
solar abundance composition for cosmic rays accelerated in these
regions and adopt the solar element abundance reported in
\citet{Lodders09} in the following calculation. For simplicity, we
attribute elements with $6<Z<10$ to oxygen, with $11\leq Z<16$ to
silicon, with $17\leq Z<23$ to calcium, and with $23\leq Z \leq 26$
to iron, and neglect heavier elements. Then the relative mass
abundance is $M_{\rm H}: M_{\rm He}: M_{\rm O}: M_{\rm Si}: M_{\rm
Ca}: M_{\rm Fe}\approx 60: 24: 1: 0.16: 0.014: 0.12$.

Since cosmic rays of  solar   abundance  composition are dominated
by light elements such as hydrogen and helium, the condition 1) can
not be satisfied and hence the condition 2) should be taken into
consideration. We assume that the maximum and minimum energies of
particles for different species are rigidity--dependent (i.e.
$E_{\rm max}, E_{\rm min}\propto Z$), and the spectrum for nuclei
with nuclear charge $Z_i$ is described by
\begin{equation}
dN_{i}/dE=f_iN_0(E/E_0)^{-s} ~~~ E_{i,\rm min}<E<E_{i, {\rm max}},
\end{equation}
where $f_i$ is the relative number abundance of element $i$ at a
given energy E, and $N_0$ and $E_0$ are used for normalization. Above
$E_{\rm max}$ or below $E_{\rm min}$, an abrupt cutoff in the
spectrum is assumed. Thus, the relative number abundance $f_i$
relates to the mass abundance by
\begin{equation}
M_i : M_j=\int_{E_{i,\rm min}}^{E_{i,\rm max}}A_i\frac{dN_i}{dE}dE : \int_{E_{j,\rm min}}^{E_{j,\rm max}}A_j\frac{dN_j}{dE}dE= A_iZ_i^{1-s}f_i : A_jZ_j^{1-s}f_j \,\,\,\, (i,j=\rm
H,He,C,...) .
\end{equation}
For a mixed composition, the maximum number of events within
$18^{\circ}$ of Cen~A  is given by
\begin{equation}
N_{\rm CenA, mix}=\frac{\sum\limits_{i} \int_{10^{19.75}\rm
eV}^{10^{19.95}\rm eV}
\zeta_i(E)\eta_i(E)\frac{dN_i}{dE}dE}{\sum\limits_{i}
\int_{10^{19.75}\rm eV}^{10^{19.95}\rm eV}
\zeta_i(E)\frac{dN_i}{dE}dE}\times N_{\rm max,mix}
\end{equation}
where {$N_{\rm max,mix}$ is the maximum event number that could be
contributed by Cen~A, as defined in \S 4, for such a mixed
composition. The value of  $N_{\rm max,mix}$ can be taken as 30,
since one can find from Figure 2 and Figure 3 that for most species
this value is roughly $\la 30$ (see discussion in Section 3). Similarly, for a mixed composition,
the average atomic mass of cosmic ray events contributed by Cen~A is
\begin{equation}
<A>=\frac{\sum\limits_{i} \int_{10^{19.75}\rm eV}^{10^{19.95}\rm eV}
A_i\zeta_i(E)\frac{dN_i}{dE}dE}{\sum\limits_{i}\int_{10^{19.75}\rm
eV}^{10^{19.95}\rm eV} \zeta_i(E)\frac{dN_i}{dE}dE}.
\end{equation}
One can see that both $N_{\rm CenA, mix}$ and $<A>$ are related with
the maximum acceleration energy $E_{i,\rm max}$ {  through the
spectrum $dN_i(E)/dE$}. As the maximum energy is rigidity-dependent,
$E_{i,\rm max}=Z_i E_{p, \rm max}$, where $E_{p, \rm max}$ is the
maximum accelerating energy for protons. Any appropriate value of
$E_{p,\rm max}$ should lead to $N_{\rm CenA,mix} \gtrsim 10$
(assuming 3.2 out of 13 events are from isotropically distributed
sources  in this region).  On the other hand, the fits to both
$<X_{\rm max}>$ and RMS$<X_{\rm max}>$ have shown that the
all-sky-averaged composition of arriving UHECRs may be nitrogen-like \citep{Taylor11}, { if only a small spread in composition exists at energies $\sim 10^{19.5}$eV}\footnote{More generally speaking, it is difficult to explain $<X_{\rm max}>$ and RMS($<X_{\rm max}>$) simultaneously by any composition because of the low value of the measured RMS($<X_{\rm max}>$). Indeed, a larger RMS($<X_{\rm max}>$) is predicted by transition models in which the composition changes from a light to a heavier composition, as is suggested by the trend in $<X_{\rm max}>$ (see e.g. \citealt{PAO11b, Ulrich09}).}.
Under the assumption that  the events contributed
by Cen~A have the same composition as the measured all--sky composition, we
conservatively set $<A>\gtrsim 10$ to be another necessary
condition.

Figure 4 shows the maximum number $N_{\rm CenA,mix}$ and the average
atomic mass $<A>$ of cosmic rays from the direction of Cen~A   as a
function of $E_{p,\rm max}$ for solar abundance composition of
cosmic rays. The red lines  represent the results for the PS model
of the GMF, while the blue ones represent the results for the J
model. One can see that in both GMF models, low $E_{p,\rm max}$
(e.g. $\la 10^{19}$eV) can be excluded, because in this case only
heavy nuclei can   be accelerated up to the required energy, while
for heavy nuclei, due to their low-rigidity,  most of them are
deflected into other directions or even can not reach the Earth. So
$E_{p, \rm max}$ should be large enough to allow at least
intermediate-mass nuclei to be able to reach $10^{19.75}$eV. On the
other hand, because of the large abundance of hydrogen or helium
elements relative to oxygen (and heavier elements), the average
chemical composition will become too light once $E_{\rm He, max} >
10^{19.75}$eV. As a consequence, we have the pale shaded region and
the dark shaded region (including the pale one) representing the
range of $E_{p,\rm max}$ in which both requirements can be satisfied
for the PS model and the J model respectively. Note that the limit
of $E_{p,\rm max}$ obtained here is consistent with the theoretical
acceleration limit of Cen~A (Lemoine \& Waxman 2009; Piran 2010),
although the acceleration ability of Cen~A is still under much
debate (Rieger \& Aharonian 2009; Pe'er \& Loeb 2011).

{Lower-energy cosmic rays that have the same rigidity ($E/Z$) will
travel along the same paths. The  chemical composition of cosmic
rays  can be constrained by requiring that they do not produce
excess in cosmic-ray flux  in the direction of Cen~A and  in the
all-sky flux at corresponding energies. As shown above, in the
viable range of $E_{p,\rm max}$ obtained, the observed excess events
from the direction of Cen A are very likely to be dominated by
intermediate mass nuclei with $6\la Z\la 10$, so we take oxygen
nuclei as the dominant particles in the following calculation. Given
that the ratio between the number of excess cosmic rays from the direction of Cen~A
and that of the whole sky is 10: 64 in the energy range from
$10^{19.75}$~eV to $10^{19.95}$~eV, cosmic rays originating from
Cen~A  should account for a flux $F_{\rm O}\simeq 10/64F^{\rm
ob}(10^{19.75-19.95}{\rm ~eV})$ within the $18^{\circ}$ window
around  Cen~A and an all--sky flux of $F_{\rm O, all}=F_{\rm
O}/\eta_{\rm O}$  in the same energy range, where $F^{\rm
ob}(10^{19.75-19.95}{\rm ~eV})=A_{\rm exp} \int_{10^{19.75}{\rm
~eV}}^{10^{19.95}{\rm eV}}\frac{dN^{\rm ob}}{dE}dE$ is the measured
all--sky flux, $dN^{\rm ob}/dE$ is the measured differential flux,
and $A_{\rm exp}$ is the total exposure of the PAO. So particles of
species $i$ that have the same rigidity in the energy range from
$\frac{Z_i}{Z_{\rm O}}10^{19.75}$~eV to $\frac{Z_i}{Z_{\rm
O}}10^{19.95}$~eV  should produce a flux of $F_i=\frac{f_{i}}{f_{\rm
O}}(\frac{Z_i}{Z_{\rm O}})^{1-s}F_{\rm O}$ inside the $18^\circ$
window of Cen A and an all--sky flux of $F_{i,\rm
all}=\frac{f_{i}}{f_{\rm O}}(\frac{Z_i}{Z_{\rm O}})^{1-s}F_{\rm
O,all}$, respectively.

Since the search at lower energies has not revealed any significant
anisotropy signals around the Cen~A region \citep{PAO11}, the flux
$F_i$ of lower-energy cosmic rays  should not be higher than the
expected background flux, i.e.
\begin{equation}\label{con_ani}
\frac{f_i}{f_{\rm O}}\left(\frac{Z_i}{Z_{\rm O}}\right)^{1-s}F_{\rm
O}\la xF^{\rm ob}\left[\frac{Z_i}{Z_{\rm O}}(10^{19.75}-10^{19.95}){\rm ~eV}\right],
\end{equation}
where $x\simeq 0.0466$ is the fraction of the exposure of the PAO
within the $18^{\circ}$ window of Cen~A \citep{PAO11}. It
should be pointed out here that the above inequality is a
conservative constraint, since the anisotropy signal can be fairly
strong when the source counts are  comparable to the background
counts, especially if the number of background counts is large. Note that this method is only applicable to lower-energy,
lighter cosmic rays.

One can also get constraints on the cosmic ray composition by
requiring that lower-energy all--sky  flux produced by sources in
the direction of  Cen A should be lower than the measured flux in
the same energy range, i.e.
\begin{equation}\label{con_flux}
\frac{f_i}{f_{\rm O}}\left(\frac{Z_i}{Z_{\rm O}}\right)^{1-s}F_{\rm
O}/\eta_{\rm O} \la F^{\rm ob}\left[\frac{Z_i}{Z_{\rm O}}(10^{19.75}-10^{19.95}){\rm ~eV}\right].
\end{equation}
The measured cosmic ray flux in the relevant range can be fitted by
a broken power-law, given by  \citep{PAO10c}
\begin{equation}
\frac{dN_{\rm ob}}{dE}=\left\{
\begin{array}{ll}
N_{\rm b}\left(\frac{E}{E_{\rm b}}\right)^{-p_1}, E<E_{\rm b}\\
N_{\rm b}\left(\frac{E}{E_{\rm b}}\right)^{-p_2}, E>E_{\rm b}
\end{array}
\right.
\end{equation}
where $E_{\rm b}=10^{19.46}$eV is the break energy, $N_{\rm b}$ is
the normalized factor,  $p_1=2.6$ and $p_2=4.3$ are the power-law
indexes of the two segment, respectively. Then from
inequality~(\ref{con_ani}) we obtain
\begin{equation}
f_i\la f_{\rm O}\times 6.4x\left(\frac{Z_i}{Z_{\rm O}}\right)^{s-p_1}\left(\frac{1-p_2}{1-p_1}\right)\left(\frac{E_2^{1-p_1}-E_1^{1-p_1}}{E_2^{1-p_2}-E_1^{1-p_2}}\right)E_{\rm b}^{p_1-p_2} ~~~ \rm for~~lighter~~elements,
\end{equation}
and from inequality~(\ref{con_flux}) we obtain
\begin{equation}
f_i\la f_{\rm O}\times \left\{
\begin{array}{ll}
6.4\eta_{\rm O}\left(\frac{Z_i}{Z_{\rm O}}\right)^{s-p_2} & \rm for~~heavier~~elements,\\
6.4\eta_{\rm O}\left(\frac{Z_i}{Z_{\rm
O}}\right)^{s-p_1}\left(\frac{1-p_2}{1-p_1}\right)\left(\frac{E_2^{1-p_1}-E_1^{1-p_1}}{E_2^{1-p_2}-E_1^{1-p_2}}
\right)E_{\rm b}^{p_1-p_2} & \rm for~~lighter~~elements .
\end{array}
\right.
\end{equation}
Here $E_1=10^{19.75}$~eV and $E_2=10^{19.95}$~eV. One can find that
for lighter elements, which method gives a stronger constraint
depends only on the value of $x$ and $\eta_{\rm O}$. In the case
discussed in the present work, $x \ll \eta_{\rm O}$, so the
constraints from the $18^\circ$ window anisotropy are stronger. But
for heavier elements, useful constraints only come from the latter
method. Assuming $s=2$,  inequality~(11) and (12) give $f_{\rm H}\la
4.4f_{\rm O}$, $f_{\rm He}\la 2.9f_{\rm O}$, $f_{\rm Si}\la
0.72f_{\rm O}$, $f_{\rm Ca}\la 0.31f_{\rm O}$ and $f_{\rm Fe} \la
0.17f_{\rm O}$, corresponding to $M_{\rm H} \la 2.2M_{\rm O}$,
$M_{\rm He} \la 2.9M_{\rm O}$, $M_{\rm Si} \la 0.72M_{\rm O}$,
$M_{\rm Ca} \la 0.31M_{\rm O}$, $M_{\rm Fe} \la 0.18M_{\rm O}$ for
the PS model; and $f_{\rm H}\la 4.4f_{\rm O}$, $f_{\rm He}\la
2.9f_{\rm O}$, $f_{\rm Si}\la 1.2f_{\rm O}$, $f_{\rm Ca}\la
0.51f_{\rm O}$ and $f_{\rm Fe} \la 0.29f_{\rm O}$, corresponding to
$M_{\rm H} \la 2.2M_{\rm O}$, $M_{\rm He} \la 2.9M_{\rm O}$, $M_{\rm
Si} \la 1.2M_{\rm O}$, $M_{\rm Ca} \la 0.51M_{\rm O}$, $M_{\rm Fe}
\la 0.31M_{\rm O}$ for the J model. A larger power-law index $s$
will put a more stringent constraint on the relative abundance of
lighter particles. One can find that, in both GMF models,
constraints on heavier elements are easy to satisfy, i.e. the excess
in flux at higher energy induced by heavier nuclei with the same
rigidity can be avoided. But to avert a low energy excess in the
Cen~A direction made by protons and helium nuclei with the same
rigidity, a super--solar metallicity especially a super--solar
intermediate-mass element abundance is required. This result is 
consistent with the analysis from fitting $X_{\rm max}$ and 
RMS($<X_{\rm max}>)$ that measured by the PAO\citep{Shaham12}.
Such a chemical composition is not favored by  the candidate cosmic-ray accelerators
in the radio galaxy Cen~A. In other words, if we attribute the 10
excess events to Cen~A, a stronger excess would also occur at 1-10
~EeV.}  We note, however, that if the composition of these 10
clustering events turns out to be proton or helium nuclei, different
from the measured all-sky-averaged composition, then the above
conclusion does not apply \citep[for light nuclei composition, see
e.g.][and reference therein]{Fargion11}.

\section{Discussion and Conclusion}
The required chemical composition of the excess cosmic-rays in the
direction of Cen A  seems to disfavor the above mentioned candidate
sites in Cen~A. The requirement, however, can be more easily
satisfied by UHECR accelerators induced by stellar explosions since
either  the stellar wind of massive stars or the exploded ejecta can
be rich in intermediate mass nuclei. The chemical composition of WR
stellar winds is $M_{\rm He} : M_{\rm C} : M_{\rm O} : M_{X}=0.32:
0.39 : 0.25: 0.04$ \citep{Bieging90, Hucht86} where $M_X$ is the sum
of elements heavier than oxygen, which  we treat simply as silicon in the 
following calculation.
The composition of the hypernova ejecta is also rich in oxygen
nuclei. The numerical modeling of the early spectra and light curve
of SN 1998bw \citep{Nakamura01} yields a composition of $M_{\rm C}:
M_{\rm O}: M_{\rm Ne}: M_{\rm Mg}: M_{\rm Si}: M_{\rm S}: M_{\rm
Ca}: M_{\rm Fe}= 0.006: 0.71: 0.037: 0.034: 0.083: 0.041: 0.007
:0.09$. We show the dependence of $N_{\rm CenA}$ and $<A>$ on
$E_{p,\rm max}$ in the WR stellar wind and hypernova ejecta
composition scenarios respectively in Figure 5. One can see that as
long as $E_{p,\rm max}$ is larger than $10^{19}$eV, which can be
reached in the GRB and hypernova scenarios \citep{Wang07, Wang08},
both $N_{\rm CenA}$ and $<A>$ meet the requirements. It has been
also shown that the spectrum and composition of UHECRs accelerated
in the WR stellar wind or in the hypernova ejecta  are compatible
with the PAO's observations \citep {Liu12}.

Both (long) GRBs and hypernovae are generally believed to trace star
formation. Recent observations by the \textit{Hubble Space
Telescope} revealed triggered star formation occurred $\la 10$~Myr
ago in the inner filament of Cen~A \citep{Crockett12}, although
Cen~A belongs to elliptical galaxies, which usually have low star
formation rates. Possible alternative sites are  GRBs and hypernovae
occurring in the Centaurus cluster, which lies behind Cen~A at a
distance of $\sim 50$~Mpc. Since there are hundreds of galaxies in
the Centaurus cluster, the star formation rate is much higher than
that in Cen~A, and as a result, the rate of GRB or hypernova is much
higher. For a distance of $\sim 50$~Mpc, intermediate-mass nuclei
such as oxygen of $\ga 60$~EeV will suffer from photo-disintegration
during the propagation, so the chemical composition at the border of
the Galaxy after propagation may be different from that at the
sources \citep[also see the discussion in][]{Taylor11a}. Figure 6
shows the propagated spectrum of oxygen nuclei from the Centaurus
cluster. The integrated flux from $10^{19.75}$eV to $10^{19.95}$eV
are about a factor of 0.15 of  the initial one. Heavier nuclei at
higher energies also suffer from severe attenuation, while helium
nuclei and protons at lower-energies ($\sim 10^{18}\rm eV -
10^{19}$eV) are almost not affected. Taking into account such
attenuation, therefore, leads to even stronger constraints on the
element abundance of light particles at the sources, i.e.
approximately reducing to 0.15 times the value obtained in the previous 
section, leading to the new constraint of $M_{\rm
H}\la 0.33M_{\rm O}$ and $M_{\rm He}\la 0.43M_{\rm O}$. Even though
this new constraint is harder to satisfy, one finds that the
chemical compositions of WR stellar winds and hypernovae ejecta can
still satisfy such a strict requirement.

In this paper, we study the origin of the observed excess of UHECRs
in the direction of Cen~A. First, by simulating the propagation of
cosmic rays in the GMF, we find that the excess events can not be
mainly composed of heavy nuclei like iron nuclei coming from the
direction of Cen~A , because the GMF has a significant influence on
their trajectories and cause their apparent arrival direction to
severely deviate from their original directions. Also, the excess
events are not likely to be dominated by light particles such as
helium nuclei or protons as  the measurements of the elongation rate
of air showers by PAO suggest a heavy or intermediate mass
composition. We show that intermediate mass nuclei with nuclear
charge $6\la Z\la 10$  are good candidates of the main composition
of the excess events. We also show that the composition of cosmic
rays can be further constrained when the anisotropy of  low-energy
cosmic rays is considered. In order not to produce a significant
anisotropy in the direction of Cen~A at lower energies, which is not
observed by the PAO,  a low proton and helium abundance (e.g. a
supersolar metallicity) is required. None of the proposed candidate
acceleration sites in Cen~A is favored by this constraint. We find
that the cosmic-ray accelerators arising from stellar explosions
such as GRBs or hypernovae are more favorable sources of the excess
events because high abundance of intermediate mass elements could be
possible in these accelerators. Since the event rates of GRBs or
hypernovae trace the star formation rate, the star formation region
in the inner filament of Cen~A or in the Centaurus Cluster  may be
the origin of the excess events.  { However, as a word of caution,
it should be remembered that the excess is significant only a posteriori.} 
{ Furthermore} we note that due to uncertainties
in our knowledge of the hadronic model describing cosmic ray
atmospheric showers, the GMFs, the chemical environment of Cen~A and
even the acceleration mechanism of UHECRs, the conclusions made in
this paper should be regarded as model-dependent.  

\acknowledgments
We thank Gwenael Giacinti, Zhuo Li and Hasan Y\"uksel for valuable comments, Xiao-Hui Sun, Yun-Ying Jiang and Li-Gang Hou for useful discussions on the GMF, Jakub Vicha, Lu Lu and Tao Wang for helps in understanding the observations of the PAO, Feng Chen for help in parallel computation. The numerical calculations in this paper have been done on the IBM Blade cluster system in the High Performance Computing Center (HPCC) of Nanjing University.  This work is supported by the NSFC under grants 10973008 and 11033002, the 973 program under grant 2009CB824800, the program of NCET, and the Fok Ying Tung Education Foundation. LRY is grateful to all the members in the group leaded by Jin-Lin Han in NAOC for their friendly hospitality and creative atmosphere.

\clearpage

\clearpage
\begin{table}
\begin{center}
\tabletypesize{\tiny} \caption{Some useful parameters obtained in
our simulation. $\bar\zeta$ is the averaged number ratio of
particles that originate from the direction of Cen A and reach the
Earth when the GMF deflection is considered to those that reach the
Earth without considering the GMF deflection, $\bar\eta$ is the
averaged number ratio of particles that reach the Earth from a
direction within $18^\circ$ of Cen A to those that reach the Earth
from all directions,  and $N_{\rm CenA}$ is the maximum event number
of cosmic rays that are confined within $18^{\circ}$ of Cen~A out of
64 UHECR events. \label{tbl-1}}
\begin{tabular}{ccccccc}
\tableline\tableline\\
\multirow{2}{*}{Elements} & \multicolumn{3}{c}{PS model} & \multicolumn{3}{c}{J model}\\
~ & $\bar\zeta$ & $\bar\eta$ & $N_{\rm CenA}$ &  $\bar\zeta$ & $\bar\eta$ & $N_{\rm CenA}$\\
\tableline
H  & 1.00 & 0.96  & 27.8 & 0.95 & 1  & 13  \\
He & 1.03 & 0.93  & 27.0 & 0.95 & 1  & 13  \\
O  & 2.12 & 0.41  & 11.9 & 1.84 & 0.66  & 20.5\\
Si & 1.80 & 0.02  & 0.58 & 2.55 & 0.25  & 8.00 \\
Ca & 0.86 & 0.008 & 0.23 & 0.79 & 0.07  & 2.31 \\
Fe & 0.88 & 0.01  & 0.33 & 0.44 & 0.004 & 0.13\\
\tableline
\end{tabular}
\end{center}
\end{table}

\begin{figure}
\epsscale{0.5} \plotone{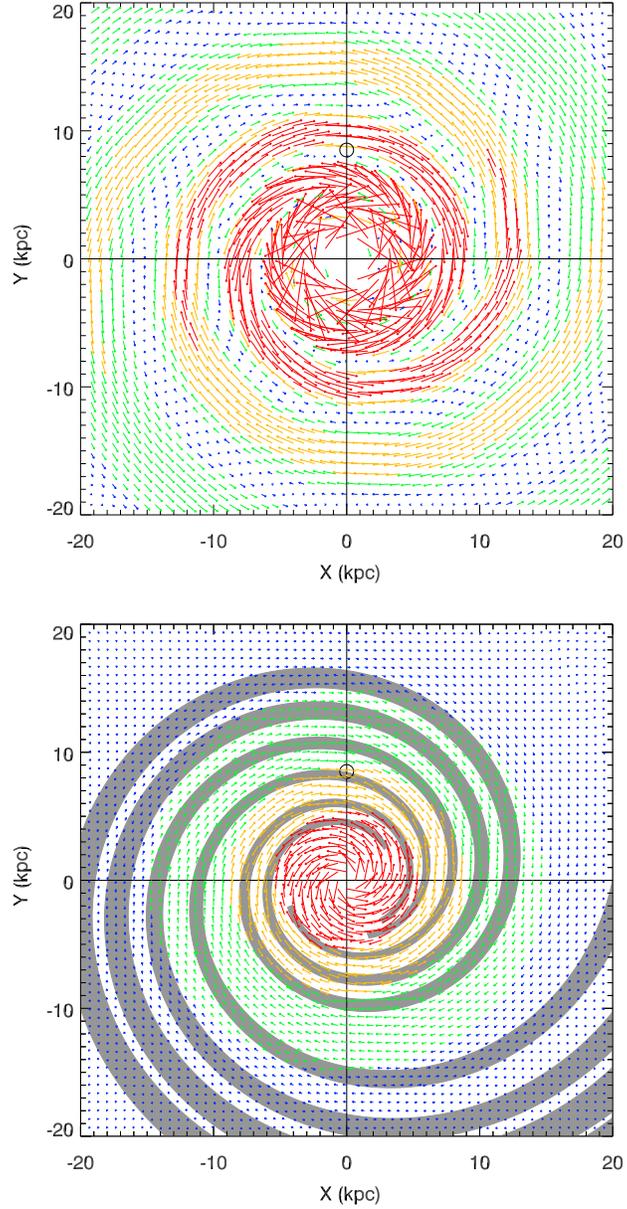}\caption{Configuration of the disk
magnetic field  in the PS model (upper panel) and the J model
(bottom panel), respectively.  The direction of arrows represents
the direction of the field while the length of arrows represents the
strength of the field. In particular, magnetic field $>3\mu$G is
colored as red, $>2\mu$G yellow, $>1\mu$G green, $<1\mu$G blue. The
open circle shows the location of the sun. For clarity, the field at
the Galactic center is not shown. { The shaded region in the bottom
panel outlines four spiral arms.}}
\end{figure}

\begin{figure}
\plotsix{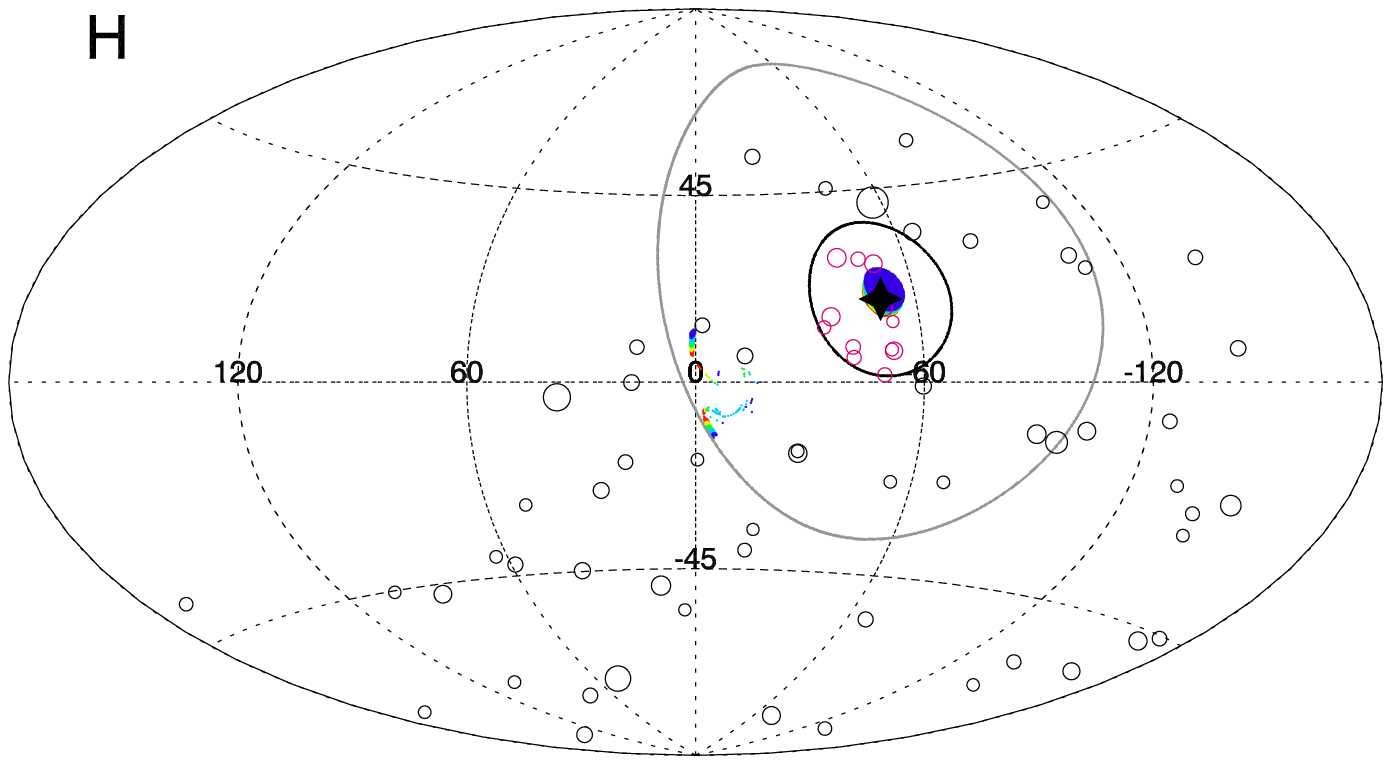}{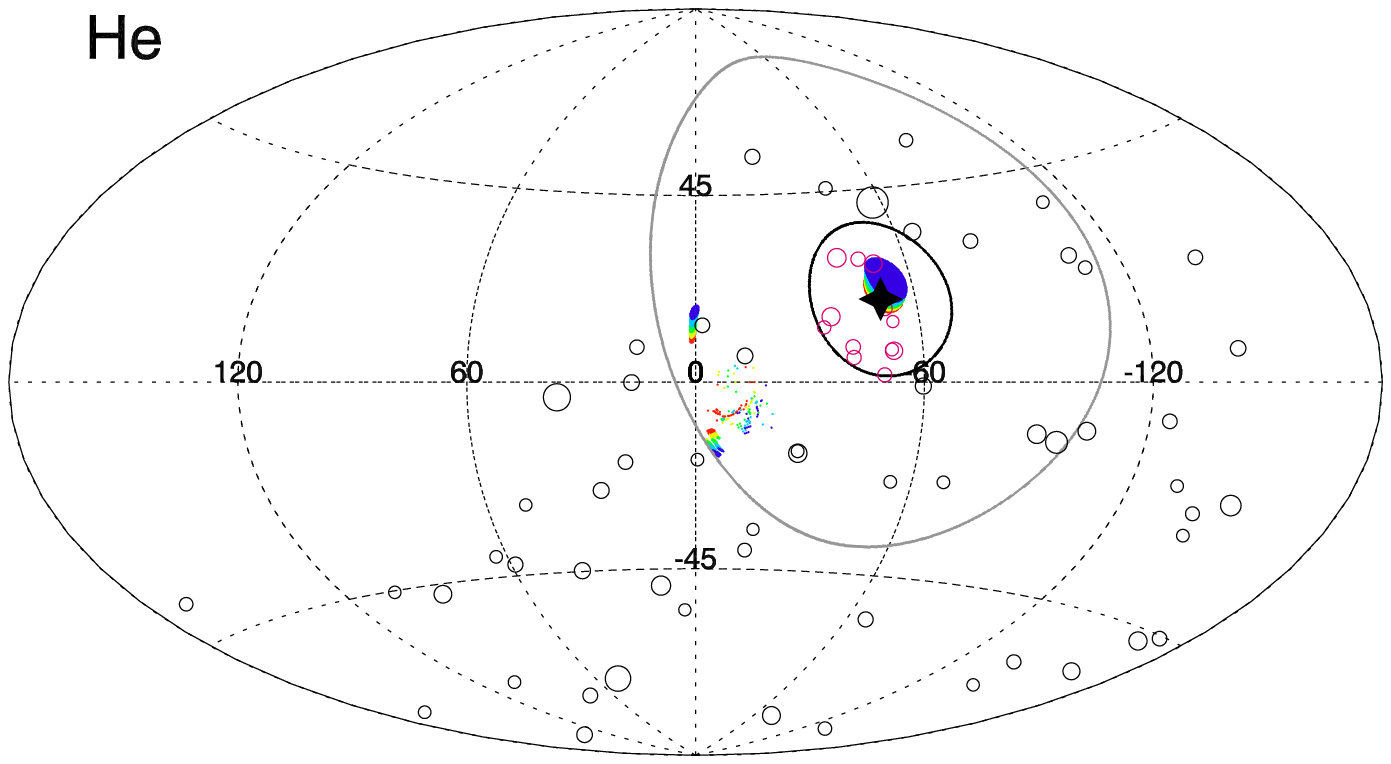}{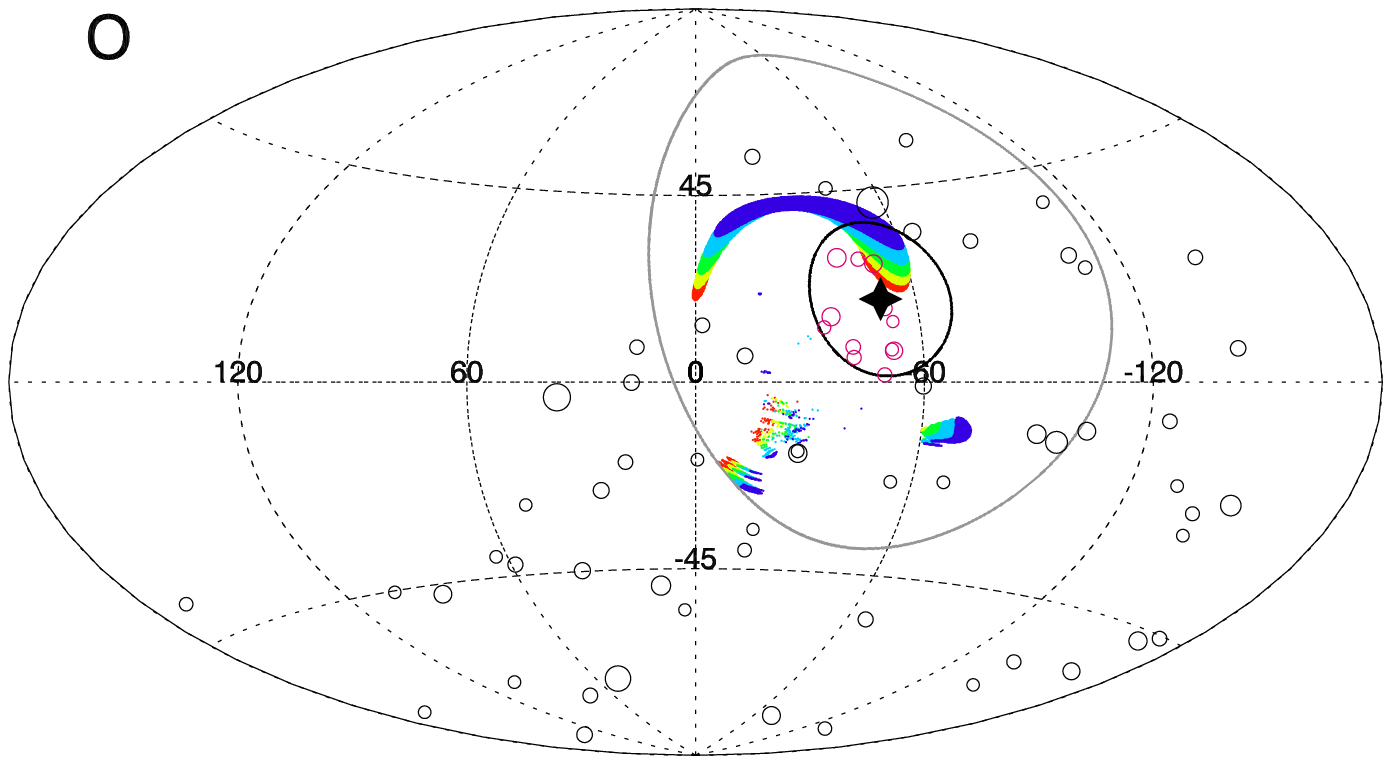}{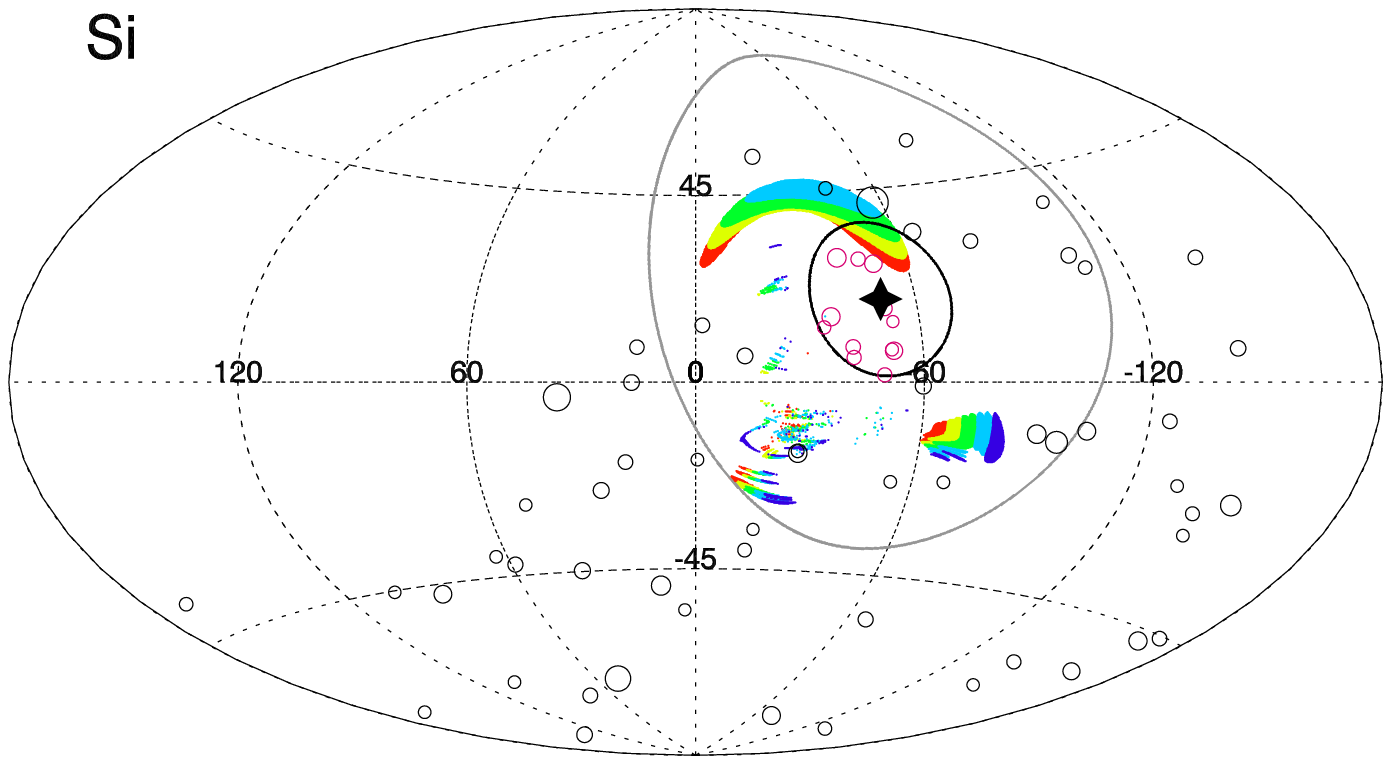}{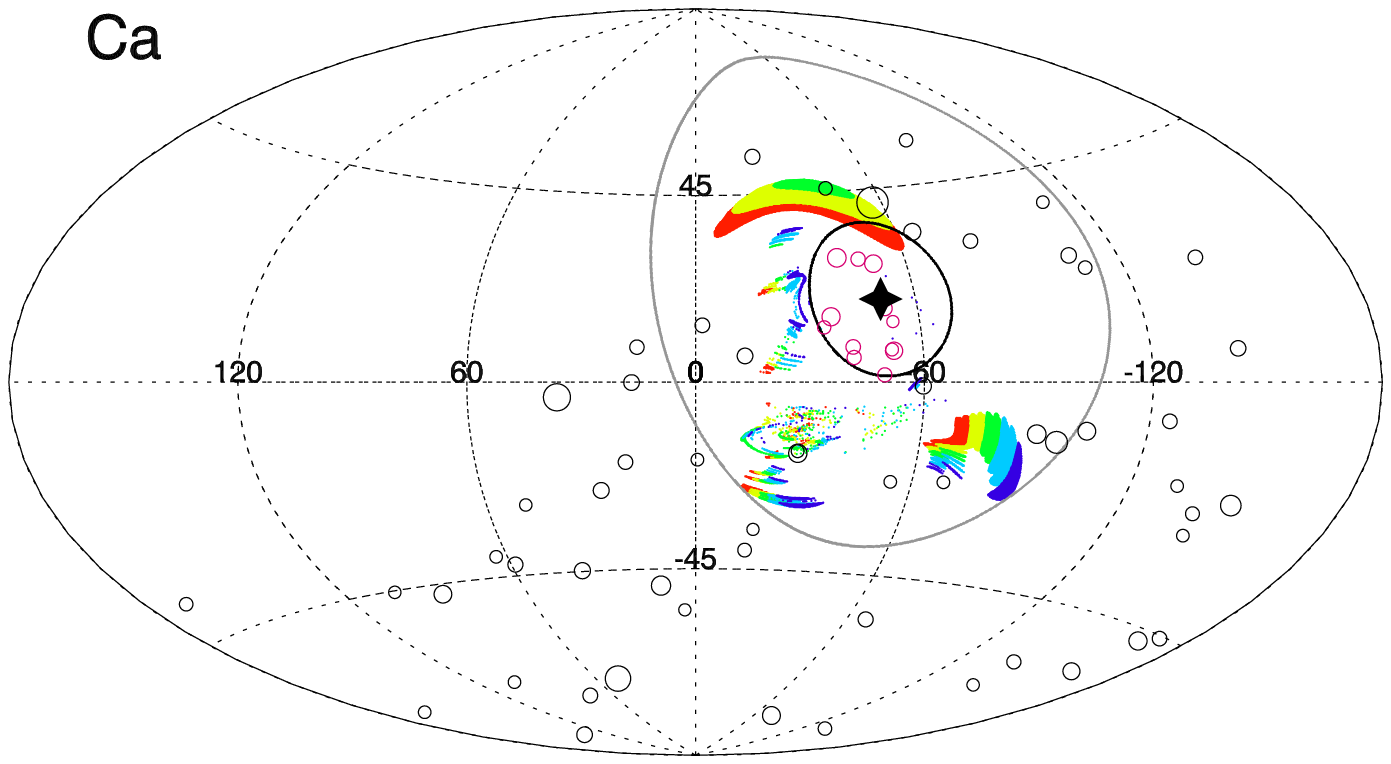}{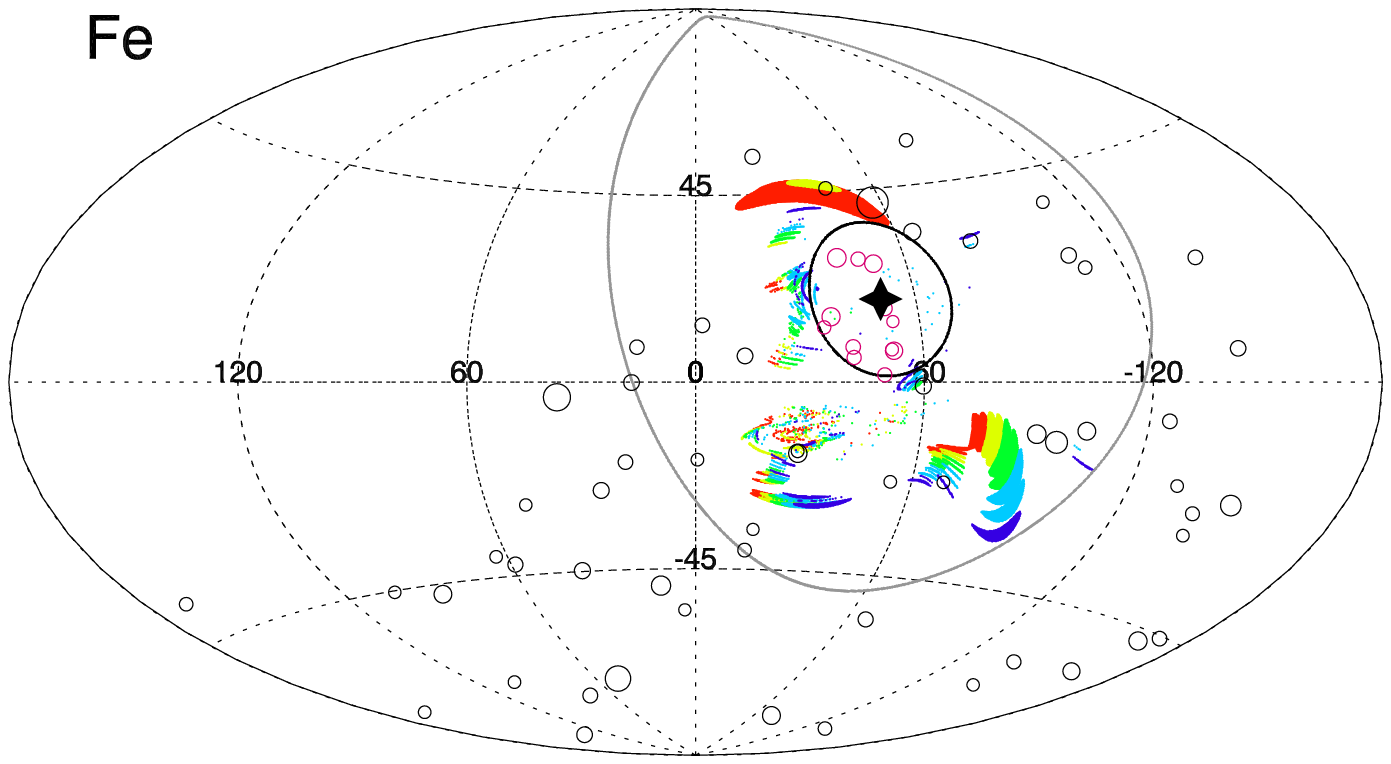}{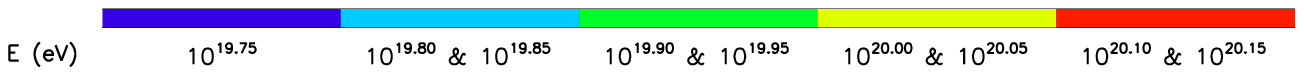}
\caption{Arrival directions  (in Galactic coordinates) of cosmic
rays originating from the direction of Cen~A in the PS model. { A total 
of $10^6$ particles are used in these simulations}. One
dot represents one particle, and different color dots/bands represent
different energies (from $10^{19.75}$eV to $10^{20.15}$eV with an increasement of $10^{0.05}$). The black filled star is the location of Cen~A
($l=309.52^{\circ}, b=19.42^{\circ}$) and the black solid curve is
the projection on the celestial sphere of a circle centered at Cen~A
with a radius of $18^{\circ}$.  The pale solid curve is the
projection of a circle centered at Cen~A with a radius equal to the
largest deflection angle of arriving cosmic rays. The small open
circles represent the observed events and the size of the circle is
proportional to the energy. The 13 { magenta} ones are those events that
are correlated with Cen~A. See text for more discussions.}
\end{figure}

\clearpage
\begin{figure}
\plotsix{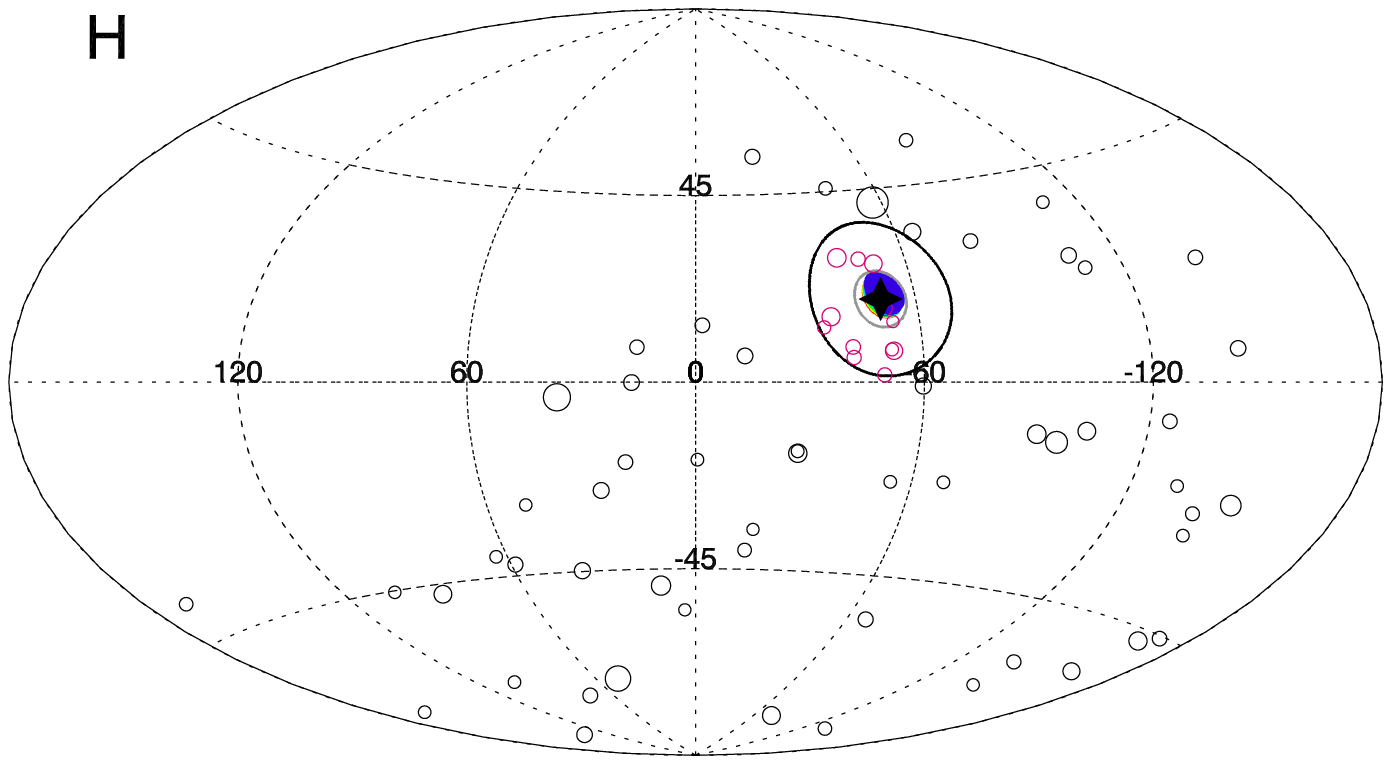}{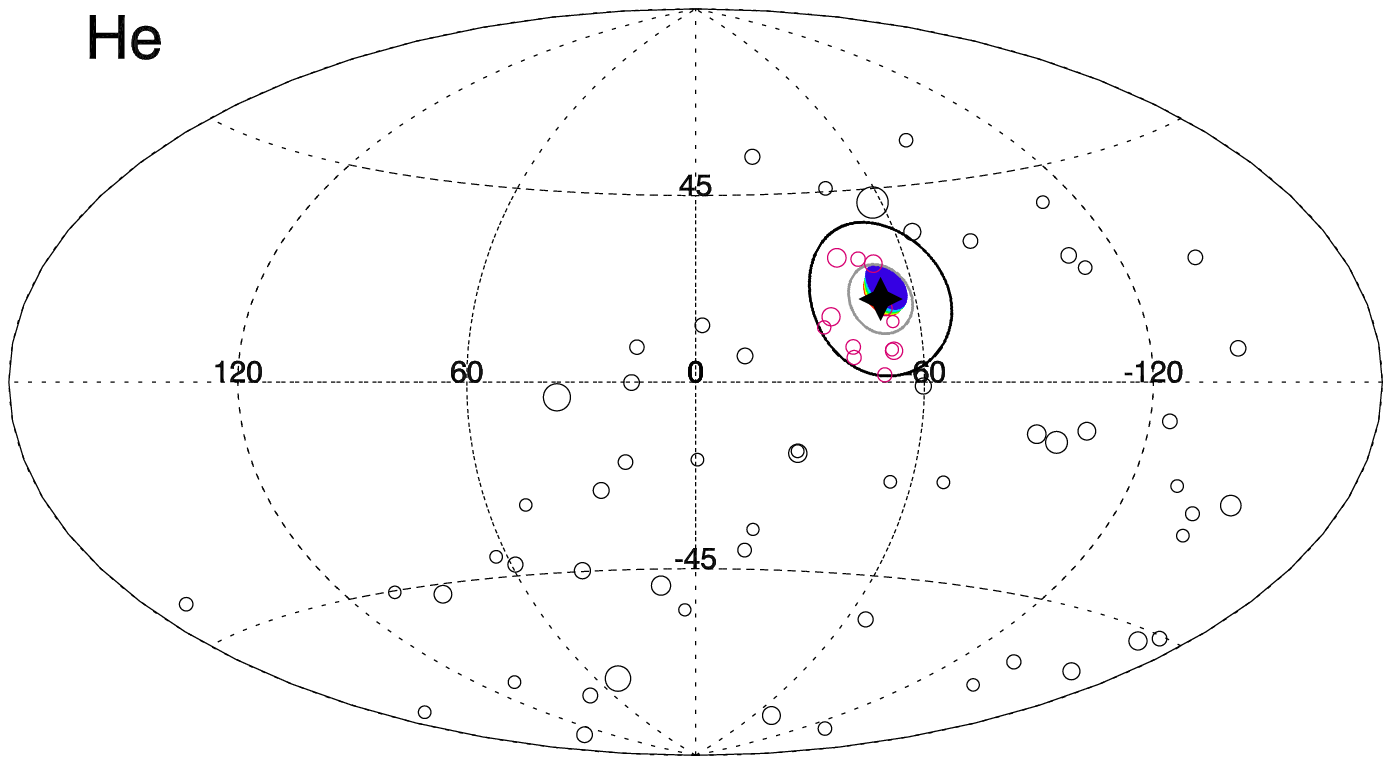}{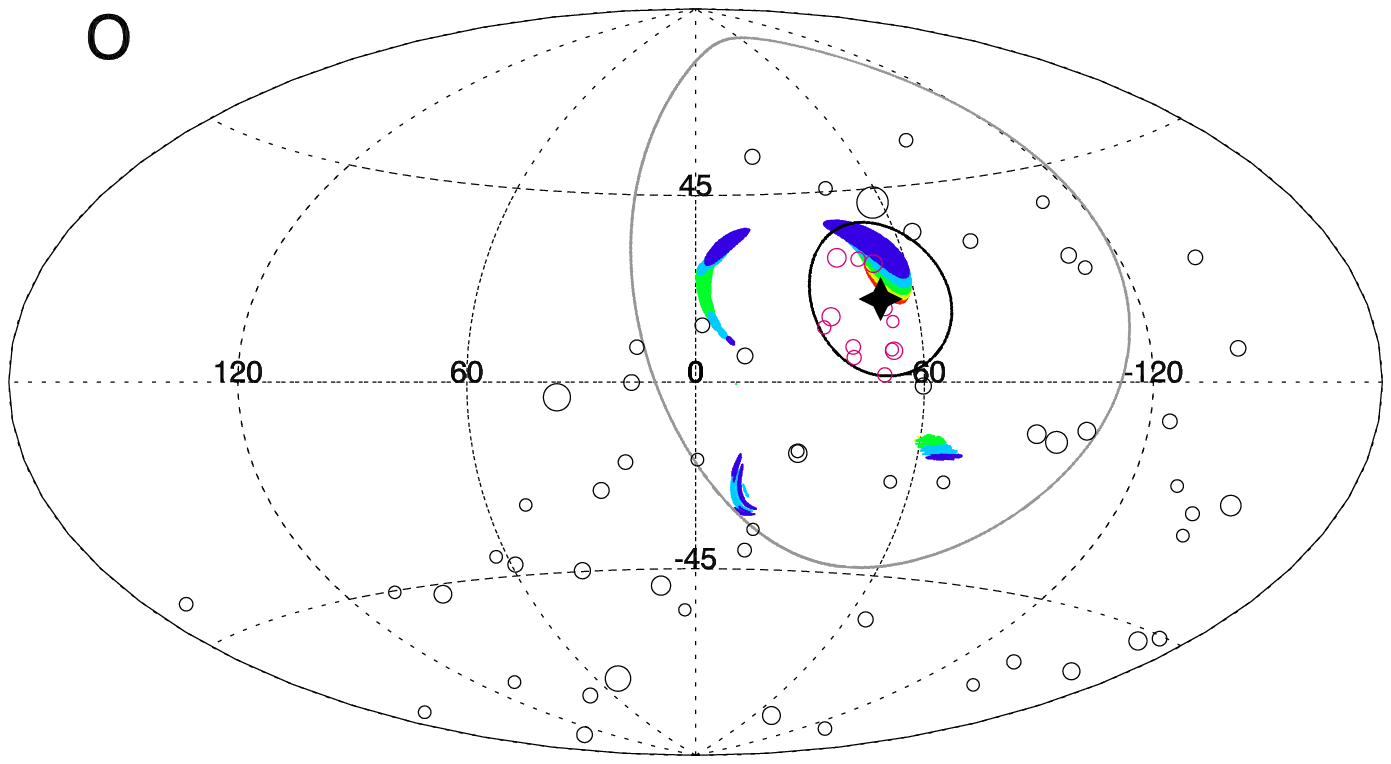}{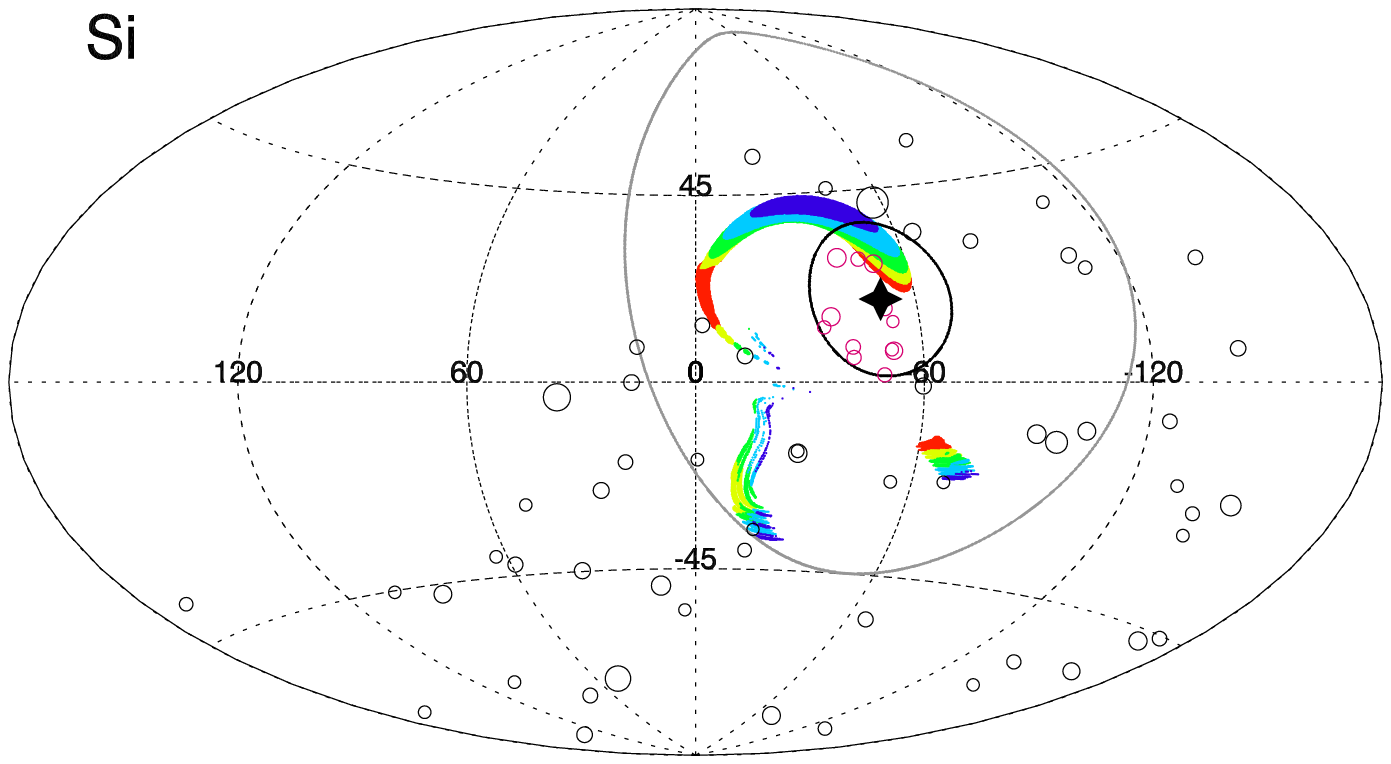}{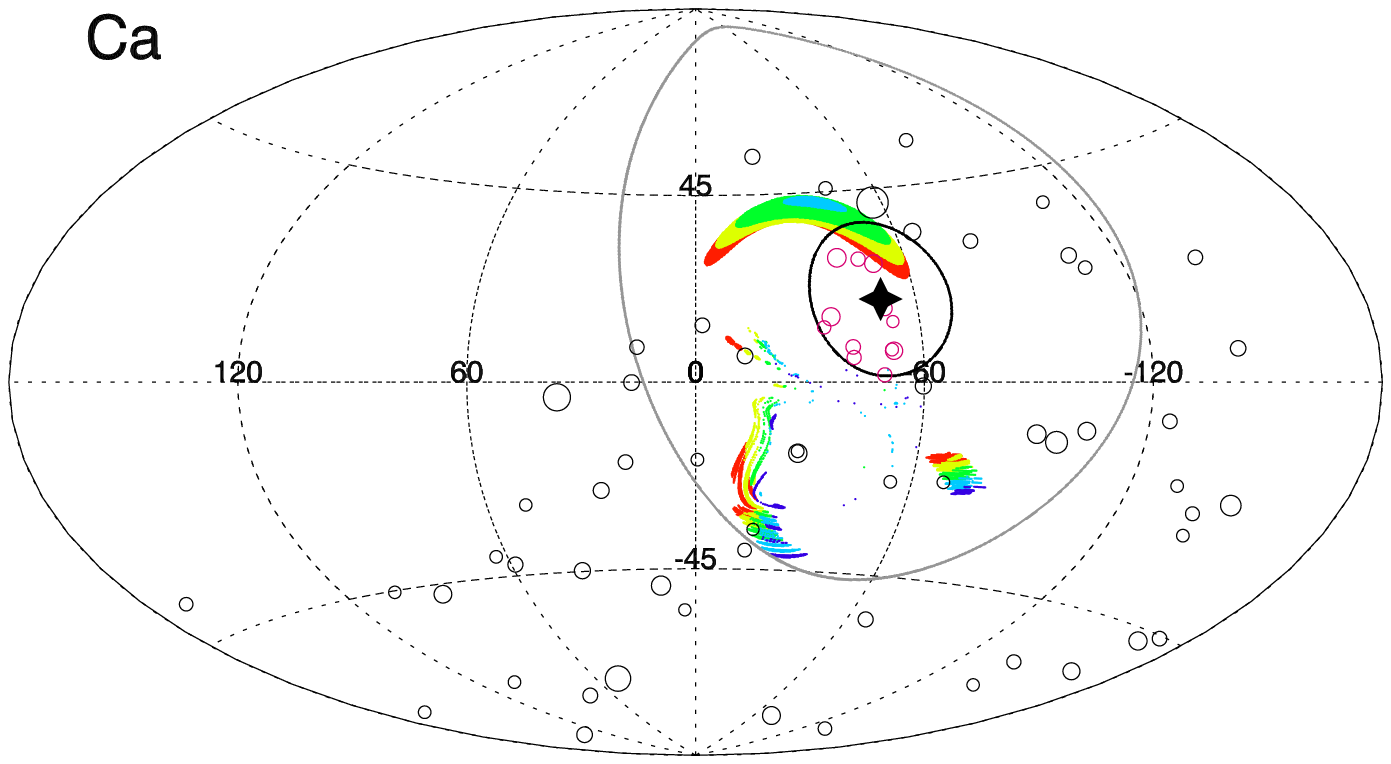}{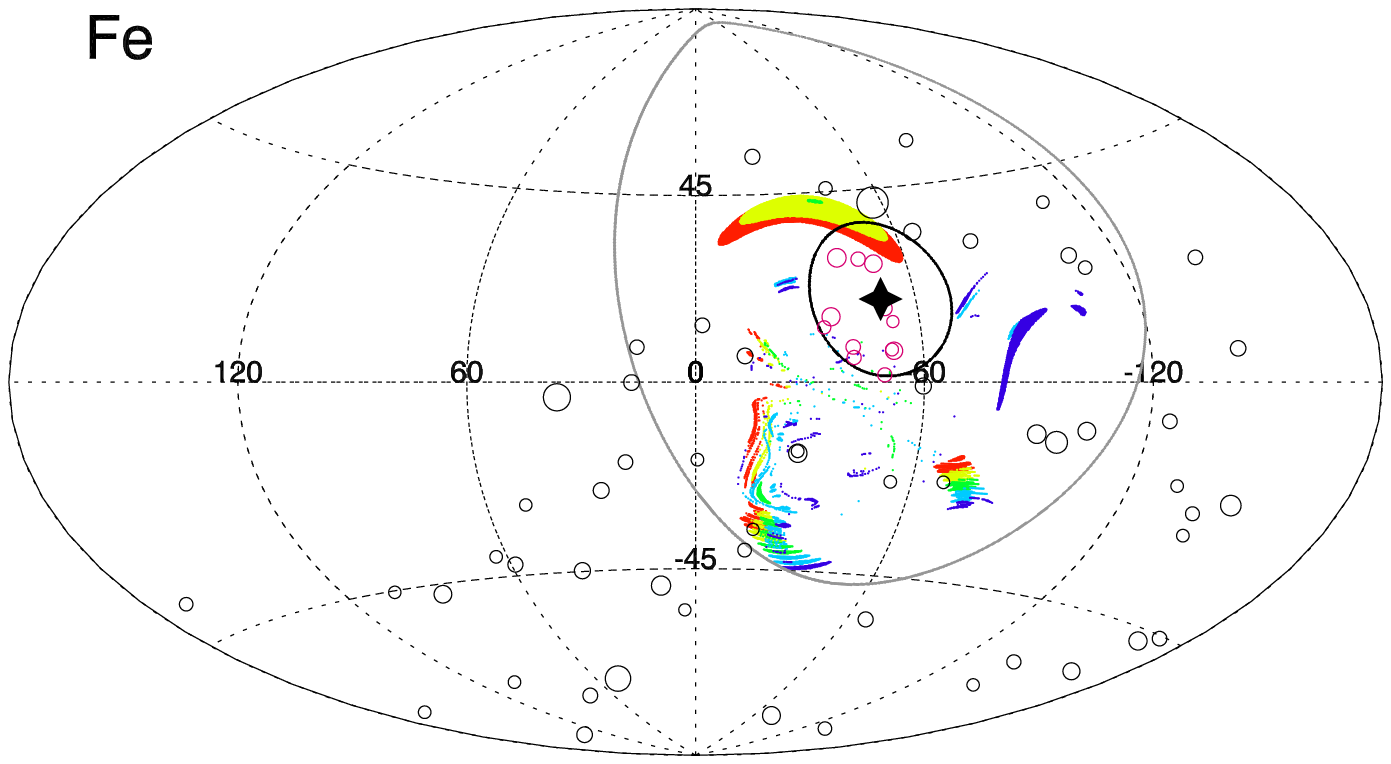}{fig2-7.eps}
\caption{The same as Figure 2 but for the case of the J model.  See text
for more discussions.}
\end{figure}

\clearpage
\epsscale{1}
\begin{figure}
\plotone{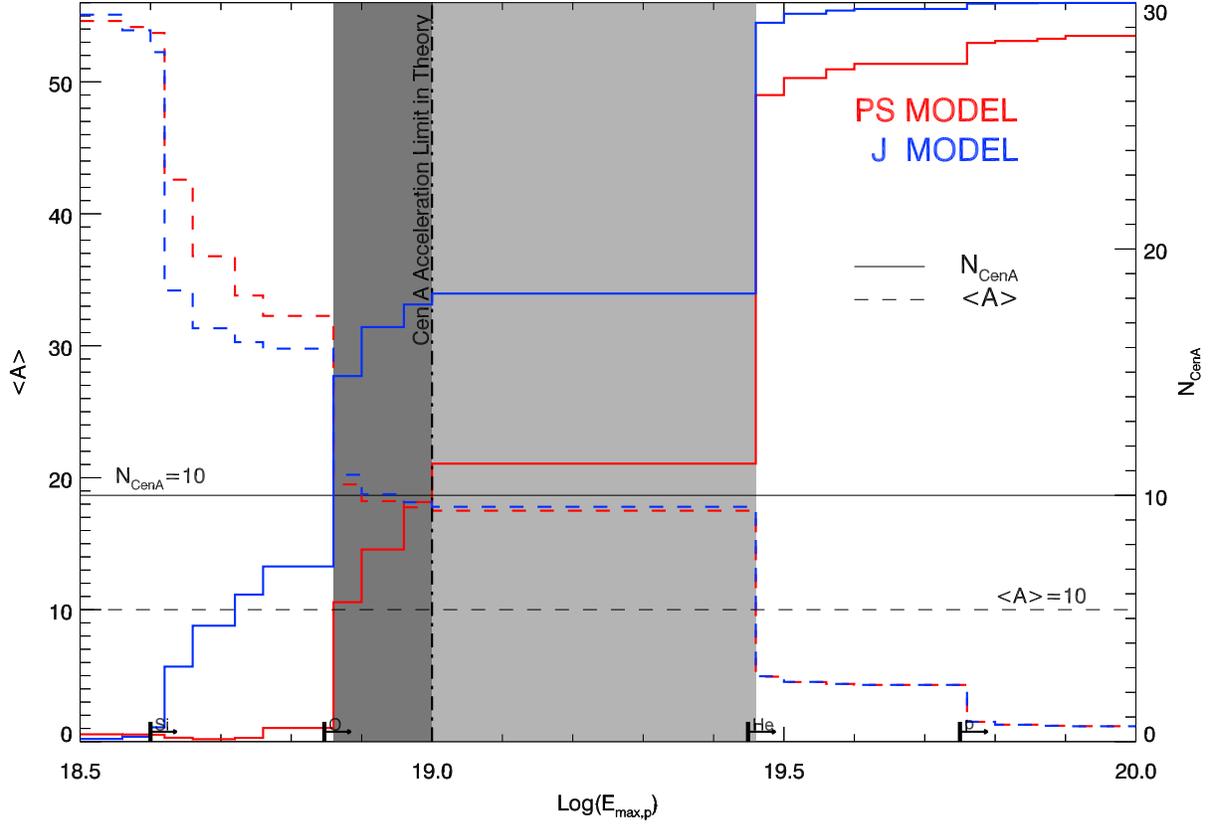}\caption{The maximum number of cosmic rays $N_{\rm
cen A}$ from the direction of Cen~A and the average atom mass $<A>$
 versus different maximum proton acceleration energy $E_{p,\rm max}$ for solar abundance composition of cosmic rays. The average
atom mass $<A>$ is shown by the dashed lines with the value indicated
by the left--handed vertical axis, while $N_{\rm cen A}$ is shown by
the solid lines with the value indicated by the right-handed
vertical axis. The corresponding values of $E_{p, \rm max}$ where
"Si", "O" and etc are indicated are the lower limits of $E_{p,\rm
max}$ for which nuclei of such elements can be accelerated to
$10^{19.75}$eV. The shaded region represents the appropriate range
of $E_{p,\rm max}$ that satisfy both $N_{\rm cen A}\ga10$ and
$<A>\ga10$. The vertical dash--dotted line shows an approximate theoretical
acceleration upper limit for protons in Cen A ($\sim 10^{19}$eV, e.g. \citealt{Hardcastle09, Lemoine09, Piran10,Gureev10}), however, which is still under much debate. See text for more discussion.}
\end{figure}

\clearpage

\begin{figure}
\epsscale{2} \plottwo{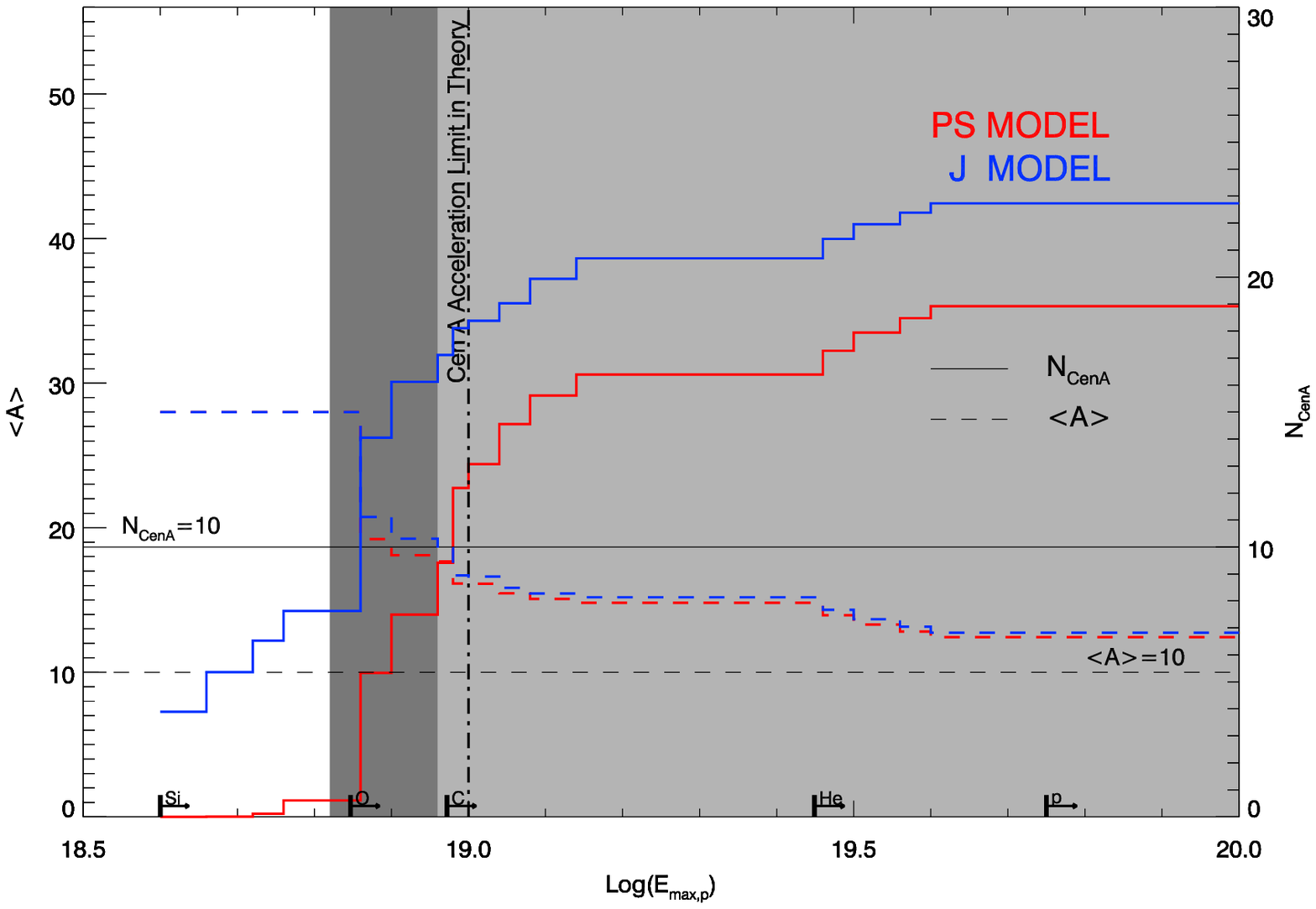}{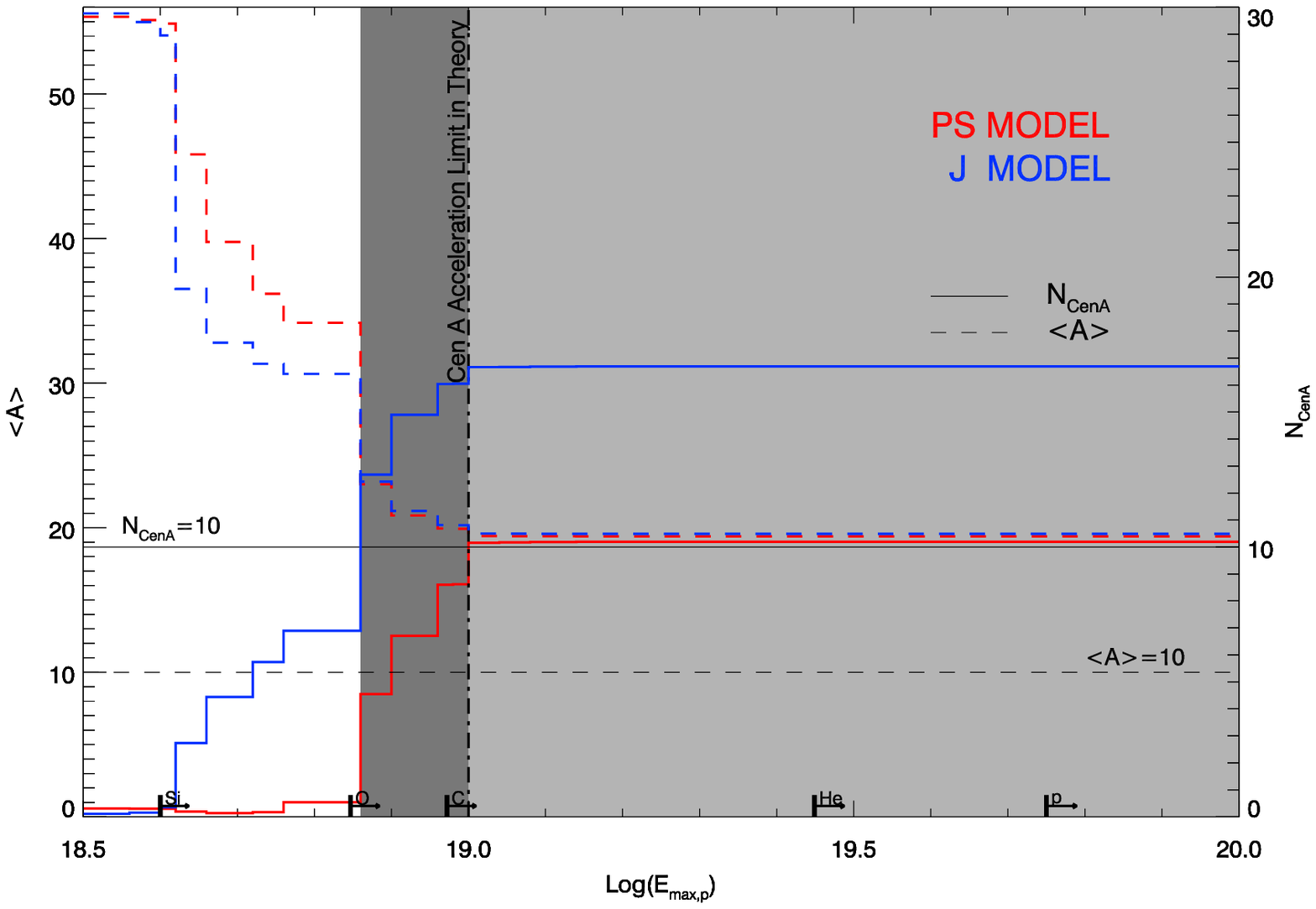}\caption{The same as
Figure 4 but for different cosmic-ray composition. The top panel
corresponds to a WR  stellar wind composition and
the bottom panel corresponds to the composition of hypernova ejecta.
See text for more discussions.}
\end{figure}

\begin{figure}
\epsscale{1} \plotone{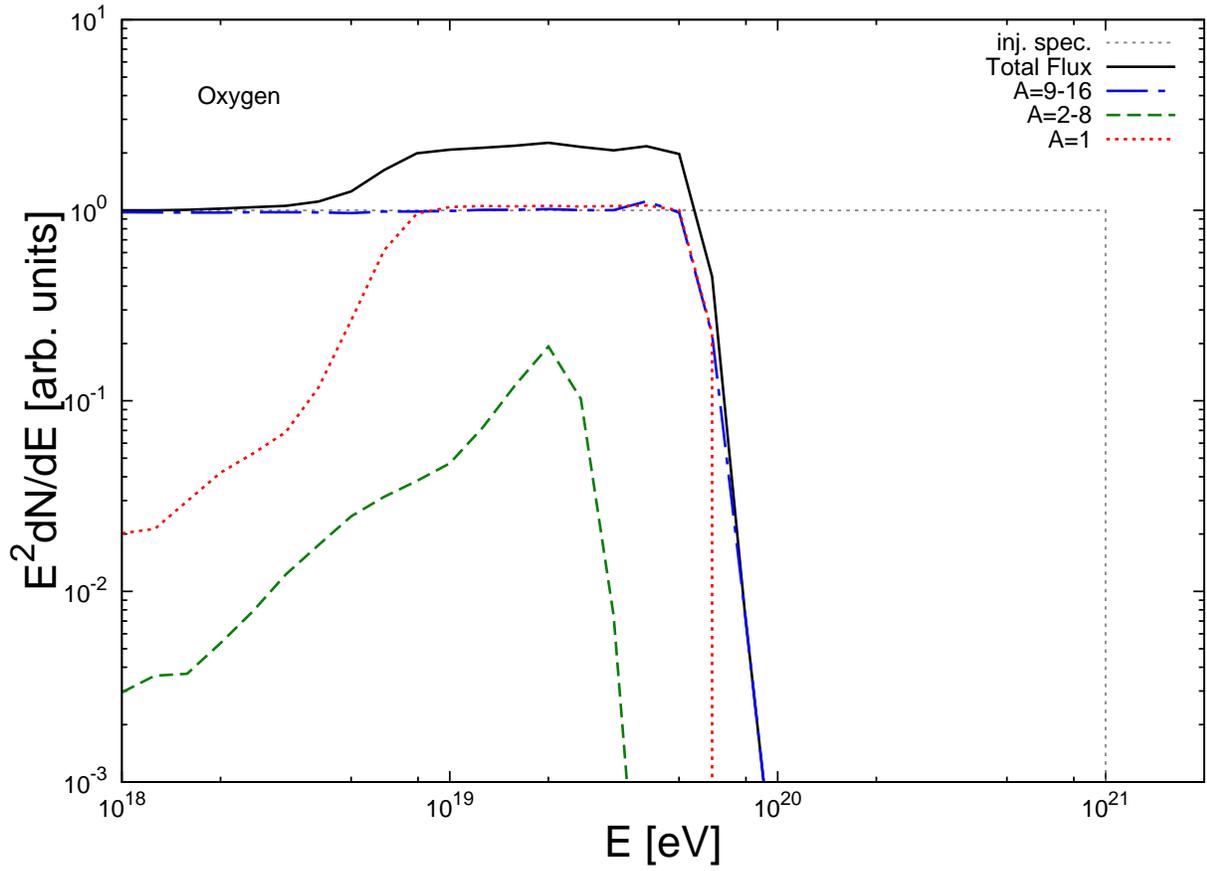}\caption{The propagated spectrum of
oxygen nuclei after traveling a distance of 50~Mpc. The initial
spectrum is set to be a power-law spectrum with index of -2 and with
an abrupt cutoff at$10^{21}$eV. }
\end{figure}

\end{document}